\documentclass[runningheads]{llncs}

\usepackage[mobile]{eccv}

\usepackage{eccvabbrv}

\usepackage{graphicx}
\usepackage{booktabs}

\usepackage[accsupp]{axessibility}  %

\usepackage{hyperref}

\usepackage{orcidlink}

\usepackage{fp}
\usepackage{expl3}[2012-07-08]
\ExplSyntaxOn
\ExplSyntaxOff

\usepackage{amsmath,amsfonts,bm}

\def\x{{x}}

\def\xi{{\x_i}}

\newcommand{\reffig}[1]{Figure~\ref{fig:#1}}
\newcommand{\refsec}[1]{Section~\ref{sec:#1}}

\newcommand{\reftbl}[1]{Table~\ref{tbl:#1}}

\newcommand{\lblfig}[1]{\label{fig:#1}}
\newcommand{\lblsec}[1]{\label{sec:#1}}

\newcommand{\lbltbl}[1]{\label{tbl:#1}}

\newcommand{\ignorethis}[1]{}
\newcommand{\myparagraph}[1]{\vspace{-2pt} \smallskip \noindent \textbf{#1}}

\def\eqref#1{equation~\ref{#1}}

\def\1{\bm{1}}

\DeclareMathAlphabet{\mathsfit}{\encodingdefault}{\sfdefault}{m}{sl}
\SetMathAlphabet{\mathsfit}{bold}{\encodingdefault}{\sfdefault}{bx}{n}

\newcommand{\ignore}[1]{}

\makeatletter
\DeclareRobustCommand\onedot{\futurelet\@let@token\@onedot}
\def\@onedot{\ifx\@let@token.\else.\null\fi\xspace}

\makeatother

\definecolor{MyDarkBlue}{rgb}{0,0.08,1}
\definecolor{MyDarkGreen}{rgb}{0.02,0.6,0.02}
\definecolor{MyDarkRed}{rgb}{0.8,0.02,0.02}
\definecolor{MyDarkOrange}{rgb}{0.40,0.2,0.02}
\definecolor{MyPurple}{RGB}{111,0,255}
\definecolor{MyRed}{rgb}{1.0,0.0,0.0}
\definecolor{MyGold}{rgb}{0.75,0.6,0.12}
\definecolor{MyDarkgray}{rgb}{0.66, 0.66, 0.66}
\definecolor{myorange}{RGB}{255,69,0}

\usepackage{multirow}
\usepackage{bm}

\title{FlashTex: Fast Relightable Mesh Texturing \\ with LightControlNet}

\author{Kangle Deng\inst{2}\thanks{Work done when interning at Roblox.} \and
Timothy Omernick\inst{1} \and
Alexander Weiss\inst{1} \and
Deva Ramanan\inst{2} \and
Jun-Yan Zhu\inst{2} \and
Tinghui Zhou\inst{1} \and
Maneesh Agrawala\inst{1,3}
}

\institute{Roblox 
\and Carnegie Mellon University 
\and Stanford University}

\authorrunning{K.~Deng et al.}
\titlerunning{FlashTex: Fast Relightable Mesh Texturing with LightControlNet}

\begin{document}

\maketitle

\begin{abstract}
Manually creating textures for 3D meshes is time-consuming, even for expert visual content creators. We propose a fast approach for automatically texturing an input 3D mesh based on a user-provided text prompt. Importantly, our approach disentangles lighting from surface material/reflectance in the resulting texture so that the mesh can be properly relit and rendered in any lighting environment. We introduce LightControlNet, a new text-to-image model
based on the ControlNet architecture, 
which allows the specification of the desired lighting as a conditioning image to the model. Our text-to-texture pipeline then constructs the texture in two stages. The first stage produces a sparse set of visually consistent reference views of the mesh using LightControlNet. The second stage applies a texture optimization based on Score Distillation Sampling (SDS) that works with LightControlNet to increase the texture quality while disentangling surface material from lighting.  Our algorithm is significantly faster than previous text-to-texture methods, while producing high-quality and relightable textures.

\end{abstract}

\section{Introduction}

Creating high-quality textures for 3D meshes is crucial across industries such as gaming, film, animation, AR/VR, and industrial design. Traditional mesh texturing tools are labor-intensive, and require extensive training in visual design. As the demand for immersive 3D content continues to surge, there is a pressing need to streamline and automate the mesh texturing process (\reffig{teaser}). %

In the past year, significant progress in text-to-image diffusion models~\cite{rombach2021stablediff,saharia2022imagen,ramesh2022dalle2} %
has created a paradigm shift in how artists create images. These models allow anyone who can describe an image in a text prompt to generate a corresponding picture.
More recently, researchers have proposed techniques for leveraging such 2D diffusion models for automatically generating textures for an input 3D mesh based on a user-specified text prompt\,\cite{chen2023text2tex,richardson2023texture,metzer2022latent,Chen_2023fantasia3D}. But these methods suffer from three significant limitations 
that restrict their wide-spread adoption in commercial applications: (1) slow generation speed (taking tens of minutes per texture), (2) potential visual artifacts (e.g., seams, blurriness, lack of details), and (3) baked-in lighting causing visual inconsistency in new lighting environments (\reffig{bakedInLighting}).  
While some recent methods address one or two of these issues, none adequately address all three.

\begin{center}
    \centering 
     
    \includegraphics[width=0.96\linewidth]{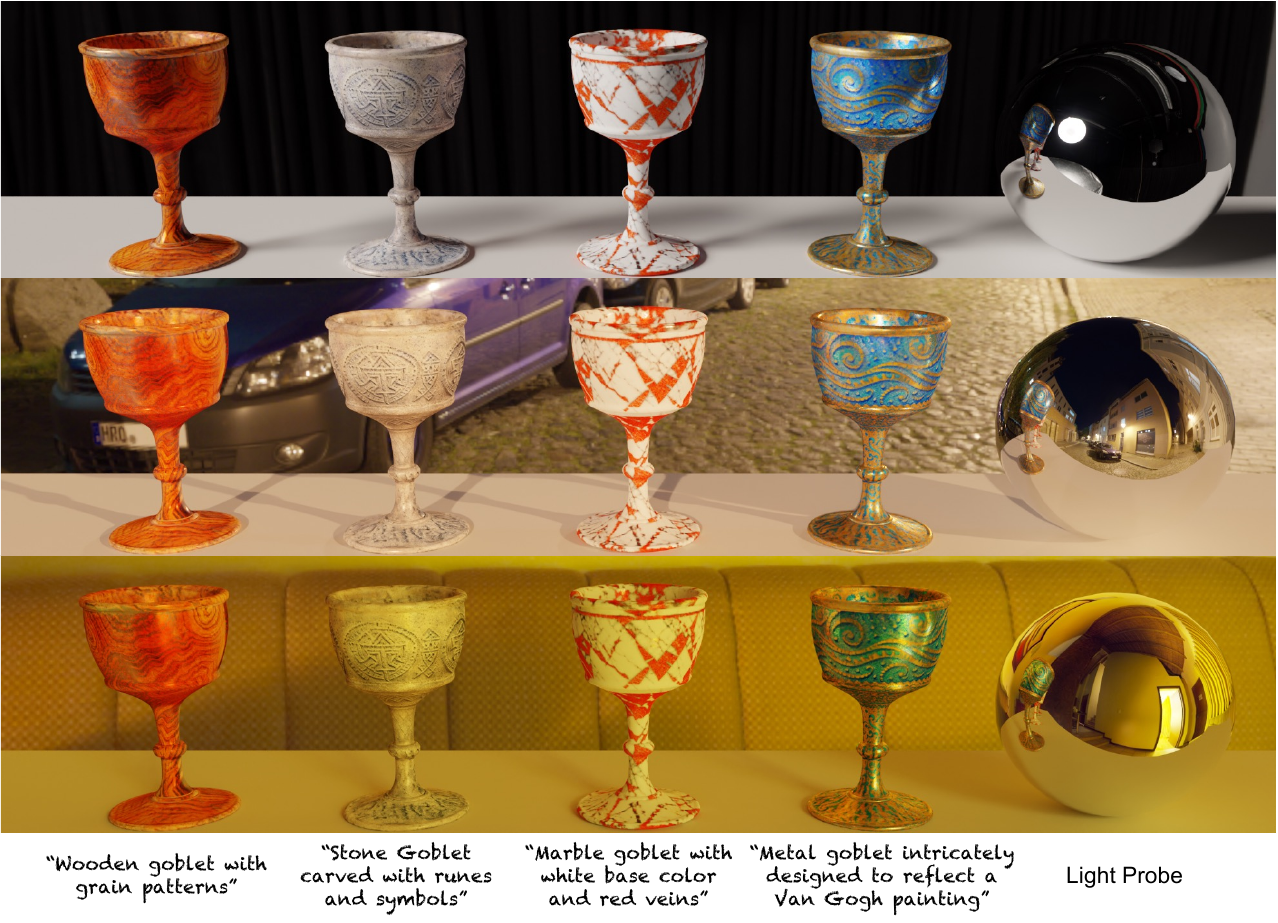}
    \captionof{figure}{We propose an efficient approach for texturing an input 3D mesh given a user-provided text prompt. Our generated texture can be relit properly in different lighting environments.  The light probe shows the varied lighting environment. We suggest the readers check our video results of rotating lighting at our \href{https://flashtex.github.io/}{website}.
    } 
    \lblfig{teaser}
\end{center}

In this work, we propose an efficient approach for texturing an input 3D mesh based on a user-provided text prompt that disentangles the lighting from surface material/reflectance to enable relighting (\reffig{teaser}).
Our method introduces {\bf LightControlNet}, an illumination-aware text-to-image diffusion model based on the ControlNet\,\cite{zhang2023controlnet} architecture, which allows specification of the desired lighting as a conditioning image for the diffusion model. Our text-to-texture pipeline uses LightControlNet to generate relightable textures in two stages. In stage 1, we use {\bf multi-view visual prompting} in combination with 
the LightControlNet to produce visually consistent reference views of the 3D mesh for a small set of viewpoints. In stage 2, we perform a new {\bf texture optimization} procedure that uses the reference views from stage 1 as guidance, and 
extends Score Distillation Sampling (SDS)\,\cite{poole2022dreamfusion} to work with LightControlNet. This allows us to increase the texture quality while disentangling the lighting from surface material/reflectance. We show that the guidance from the reference views allows our optimization to generate textures with over 10x speed-up than previous SDS-based relightable texture generation methods such as Fantasia3D \cite{Chen_2023fantasia3D}. %
Furthermore, our experiments show that the quality of our textures is generally better than those of existing baselines in terms of FID, KID, and user study.

\begin{figure*}[!t]
   \centering
   \includegraphics[width=\linewidth]{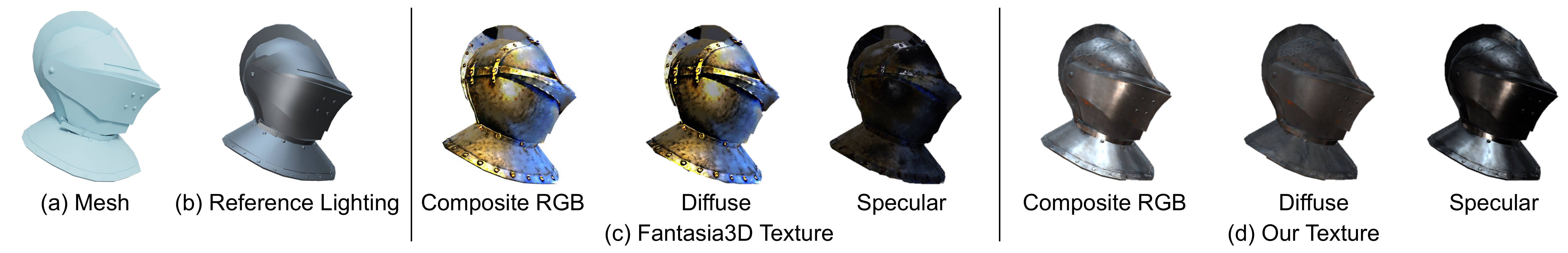}
   \caption{Given a 3D mesh of a helmet (a) and a lighting environment $L$,  
   the reference rendering (b) depicts the ``correct'' highlights on the mesh due to $L$, by treating its surface reflectance as half-metal and half-smooth with a gray diffuse color. (c) The texture generated by the leading method Fantasia3D\,\cite{Chen_2023fantasia3D} is not properly relit as Fantasia3D bakes most of the lighting into the diffuse texture for the mesh and does not capture the bright highlights in the specular texture.  (d) In contrast, our pipeline disentangles lighting from material,   better capturing the diffuse and specular components of the metal helmet in this environment. Text prompt: ``A medieval steel helmet.''
   }
   \lblfig{bakedInLighting}
\end{figure*}

\section{Related Work}
\myparagraph{Text-to-Image generation.} Recent years have seen significant advancements in text-to-image generation empowered by diffusion models \cite{ramesh2022dalle2,saharia2022imagen,rombach2021stablediff}. Stable Diffusion~\cite{rombach2021stablediff}, for example, trains a latent diffusion model (LDM) on the latent space rather than pixel space, delivering highly impressive results with affordable computational costs. 
Further extending the scope of text-based diffusion models, works such as GLIGEN~\cite{li2023gligen}, PITI~\cite{wang2022pretraining},  T2IAdapter~\cite{mou2023t2i}, and ControlNet~\cite{zhang2023controlnet} incorporate spatial conditioning inputs (e.g., depth maps, normal maps, edge maps, etc.) to enable localized control over the composition of the result. 
Beyond their power in image generation, these 2D diffusion models, trained on large-scale text-image paired datasets, also contribute valuable priors to various other tasks such as image editing~\cite{hertz2022prompt,meng2022sdedit}, 3D generation~\cite{poole2022dreamfusion,raj2023dreambooth3d}, and 3D editing~\cite{instructnerf2023,kobayashi2022decomposing,wang2022clip,zhuang2023dreameditor}.

\myparagraph{Text-to-3D synthesis.} The success of text-to-image synthesis has sparked considerable interest in its 3D counterpart. Some approaches~\cite{shue2023triplane_diffusion, zhou2021point_voxel_diffusion, li2023diffusion_sdf, nam20223d_ldm} train a text-conditioned 3D generative model akin to 2D models, while others employ 2D priors from pre-trained diffusion models for optimization~\cite{poole2022dreamfusion, wang2023sjc, lin2023magic3d, Chen_2023fantasia3D, metzer2022latent, wang2023prolificdreamer, sun2023dreamcraft3d, sweetdreamer} and multi-view synthesis~\cite{liu2023zero1to3, shi2023MVDream}. For instance, DreamFusion~\cite{poole2022dreamfusion} and Score Jacobian Chaining~\cite{wang2023sjc} were the first to propose Score Distillation Sampling to optimize a 3D representation using 2D diffusion model gradients. 
Zero-1-to-3~\cite{liu2023zero1to3} synthesizes novel views using a pose-conditioned 2D diffusion model. 
Yet, these methods often produce blurry, low-frequency textures that bake lighting into surface reflectance.
Fantasia3D~\cite{Chen_2023fantasia3D} can generate more realistic textures by incorporating physics-based materials. However, the resulting materials remain entangled with lighting, making it difficult to relight the textured object in %
a new lighting environment. In contrast, our method effectively disentangles the lighting and surface reflectance texture. 
Concurrent to our work, MATLABER~\cite{xu2023matlaber} aims to recover material information in text-to-3D generation using a material autoencoder. Our method, however, differs in approach and improves efficiency.

\begin{figure*}[t!]
    \centering
    \includegraphics[width=\linewidth]{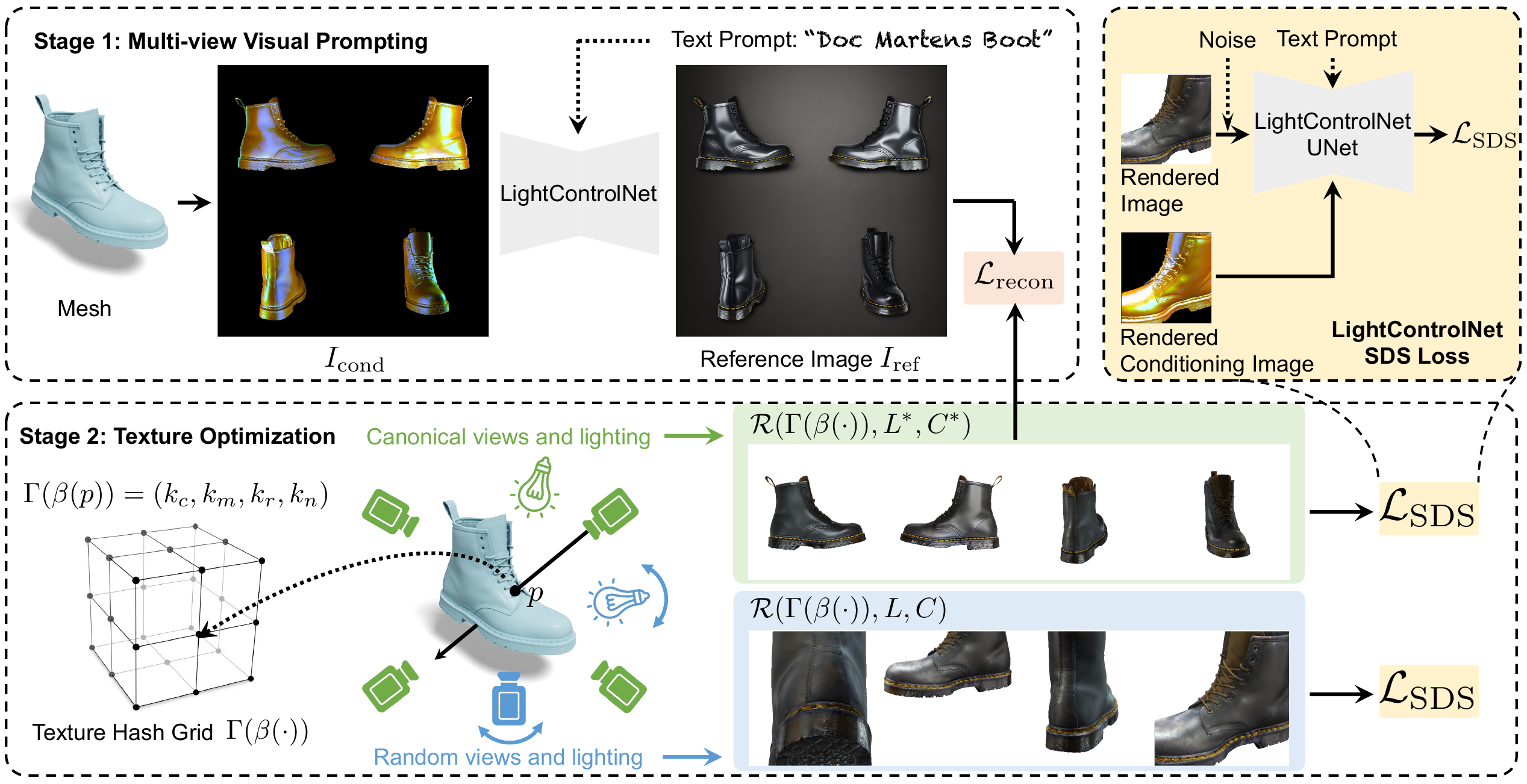}
    \caption{\textbf{Text-to-Texture pipeline.} Our method efficiently synthesizes relightable textures given a 3D mesh and text prompt. In stage 1 (top left), we use {\bf \em multi-view visual prompting} with our LightControlNet to generate four visually consistent canonical views of the mesh under fixed lighting, concatenated into a reference image $I_{\text{ref}}$. In stage 2, we apply a new {\bf \em texture optimization} procedure using $I_{\text{ref}}$ as guidance along with a multi-resolution hash-grid representation of the texture $\Gamma(\beta(\cdot))$.
    For each iteration,  we render two batches of images using $\Gamma(\beta(\cdot))$ -- 
    one using the canonical views and lighting of $I_{\text{ref}}$ to compute a reconstruction loss $\mathcal{L}_{\text{recon}}$ and the other using randomly sampled views and lighting to compute an SDS loss $\mathcal{L}_{\text{SDS}}$ based on LightControlNet. }
    \lblfig{pipeline}
\end{figure*}

\myparagraph{3D texture generation.} The area of 3D texture generation has evolved over time. Earlier models either directly took 3D representations as input to neural networks~\cite{bokhovkin2023mesh2tex,siddiqui2022texturify,yu2021learning} or used them as templates \cite{park2018photoshape, pavllo2021textured3dgan}. While some methods also use differentiable rendering to learn from 2D images~\cite{bokhovkin2023mesh2tex,henderson20cvpr,yu2021learning,pavllo2021textured3dgan}, the models often fail to generalize beyond the limited training categories. Closest to our work are the recent works that use pre-trained 2D diffusion models and treat texture generation as a byproduct of text-to-3D generation. Examples include Latent-Paint~\cite{metzer2022latent}, which uses Score Distillation Sampling in latent space, Text2tex~\cite{chen2023text2tex}, which leverages depth-based 2D ControlNet, and TEXTure~\cite{richardson2023texture}, which exploits both previous methods. Nonetheless, similar to recent text-to-3D models, such methods produce textures with entangled lighting effects and suffer from slow generation. On the other hand, TANGO~\cite{ChenChen2022tango}, generates material textures using a Spherical-Gaussian-based differentiable renderer, but struggles with complex texture generation.  A concurrent work, Paint3D~\cite{zeng2023paint3d}, aims to generate lighting-less textures, yet it cannot produce material-based textures like ours.

\myparagraph{Material generation.}
Bidirectional Reflection Distribution Function (BRDF) \cite{nicodemus1965brdf} is widely used for modeling surface materials in computer vision and graphics. Techniques for recovering material information from images often leverage neural networks to resolve the inherent ambiguities when applied to  a limited range of view angles  
or unknown illuminations. However, these methods often require controlled setups \cite{li2018materials} or curated datasets \cite{gao2019deep,bi2020neural,wang2021learning}, and struggle with in-the-wild images.
Meanwhile, material generation models like ControlMat~\cite{vecchio2023controlmat}, Matfuse~\cite{vecchio2023matfuse}, and Matfusion \cite{Sartor2023matfusion} use diffusion models for generating Spatially-Varying BRDF (SVBRDF) maps but limit themselves to 2D generation. In contrast, our method creates relightable materials %
for 3D meshes.

\section{Preliminaries}
\lblsec{prelims}

Our text-to-texture pipeline builds on several techniques that have been recently introduced for text-to-image diffusion models. Here, we briefly describe these prior methods and then present our pipeline in \refsec{method}.

\myparagraph{ControlNet.}
ControlNet~\cite{zhang2023controlnet} is an %
architecture designed to add spatially localized compositional controls to a text-to-image diffusion model, such as Stable Diffusion~\cite{rombach2021stablediff}, in the form of conditioning imagery (e.g., Canny edges~\cite{canny1986computational}, OpenPose keypoints~\cite{cao2017realtime}, depth images, etc.). 
In our work, where we take a 3D mesh as input, the conditioning image $I_{\text{cond}}(C)$ is a rendering of the mesh from a given camera viewpoint $C$. Then, given text prompt $y$, 
\begin{align*}
    I_{\text{out}} = \text{ControlNet}(I_{\text{cond}}(C), y),
\end{align*}
where the output image $I_{\text{out}}$ is conditioned on $y$ and $I_{\text{cond}}$. ControlNet introduces a parameter $s$ that sets the strength of the conditioning image. When $s=0$, the ControlNet simply produces an image using the underlying Stable Diffusion model, and when $s=1$, the conditioning is strongly applied.

\myparagraph{Score Distillation Sampling (SDS).} 
DreamFusion~\cite{poole2022dreamfusion} optimizes a 3D NeRF representation $\theta$~\cite{mildenhall2020nerf,barron2021mipnerf}
conditioned on text prompts 
using a pre-trained 2D text-to-image diffusion model.
A differentiable renderer $\mathcal{R}$ applied to $\theta$ with a randomly sampled camera view $C$ then generates a 2D image $x = \mathcal{R}(\theta, C)$. A small amount of noise $\epsilon \sim \mathbb{N}(0,1)$ is then added to $x$ to obtain a noisy image $x_t$.
DreamFusion leverages a diffusion model $\phi$ (Imagen~\cite{saharia2022imagen}) to provide a score function $\hat{\epsilon}_{\phi}(x_t; y, t)$, which predicts the sampled noise $\epsilon$ given the noisy image $x_t$, text prompt $y$, and noise level $t$. This score function can update the scene parameters $\theta$, using the gradient calculated by SDS:
\begin{align*}
    \nabla_{\theta} \mathcal{L}_{\text{SDS}}(\phi, x) = \mathbb{E}_{t,\epsilon}\left[ w(t) (\hat{\epsilon}_{\phi}(x_t; y, t) - \epsilon) \frac{\partial x}{\partial \theta} \right],
\end{align*}
where $w(t)$ is a weighting function. %
During each iteration, to calculate the SDS loss, we randomly choose a camera view $C$, render the NeRF $\theta$ to form an image $x$, add noise $\epsilon$ to it, and predict the noise using the diffusion model $\phi$. %
DreamFusion optimzes for 5,000 to 10,000 iterations. In our work, we introduce an illumination-aware SDS loss to optimize surface texture on a mesh to suppress inconsistency artifacts and simultaneously separate lighting from the material.

\section{Method}
\lblsec{method}

Our text-to-texture pipeline operates in two main stages to generate a relightable texture for an input 3D mesh with a text prompt (\reffig{pipeline}). 
In Stage 1, we use {\bf multi-view visual prompting} to obtain visually consistent views of the object from a small set of viewpoints, using a 2D ControlNet. Simply 
backprojecting these sparse views onto the 3D mesh could produce patches of high-quality texture but would also generate visible seams and other visual artifacts where the views do not fully match. The resulting texture would also have lighting baked-in, making it difficult to relight the textured mesh in a new lighting environment.  Therefore, in Stage 2, we apply a {\bf texture optimization} that uses a ControlNet in combination with Score Distillation Sampling (SDS)~\cite{poole2022dreamfusion} 
to mitigate such artifacts and separate lighting from the surface material properties. 
In both stages, we introduce a new illumination-aware ControlNet that allows us to specify the desired lighting as a conditioning image for an underlying text-to-image diffusion model. We call this model {\bf LightControlNet} and describe how it works in \refsec{LightControlNet}. We then detail each stage in \refsec{stage_1} and \refsec{stage_2}, respectively.

\vspace{-0.1in}
\begin{figure*}[t]
    \centering
    \includegraphics[width=\linewidth]{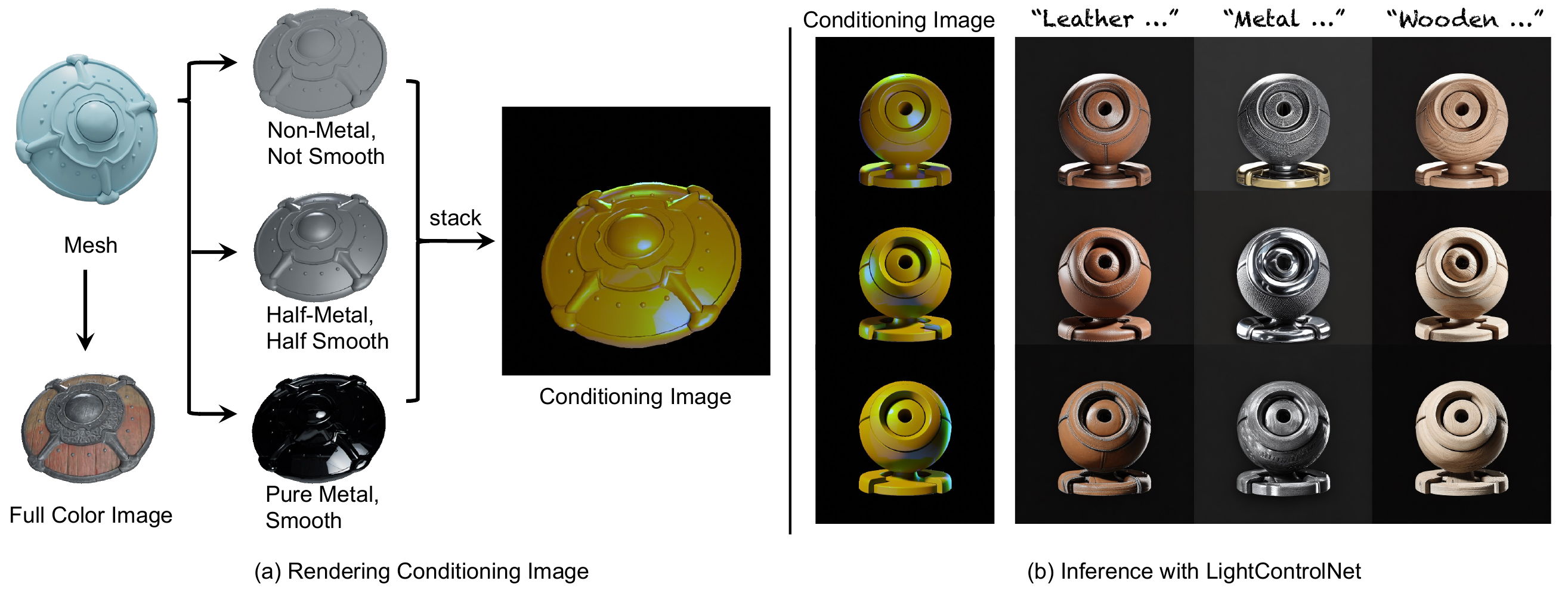}
    \caption{(a) LightControlNet requires a conditioning image that specifies desired lighting $L$ for a view $C$ of a 3D mesh. To form the conditioning image, we render the mesh with the desired $L$ and $C$ using three different materials: (1) non-metal, not smooth, (2) half-metal, half-smooth, and (3) pure metal, smooth, and then combine the renderings into a single three-channel image. (b) LightControlNet is a diffusion model conditioned on such light-conditioning images and text prompts.
}
    \lblfig{illum_control}
\end{figure*}

\subsection{LightControlNet}
\lblsec{LightControlNet}
LightControlNet adapts the ControlNet architecture to enable control over the lighting in the generated image. 
Specifically, we create a conditioning image for a 3D mesh by rendering it using three pre-defined materials and under known lighting conditions (\reffig{illum_control}). These renderings encapsulate information about the desired shape and lighting for the object, and we stack them into a three-channel conditioning image. 
We have found that setting the pre-defined materials to (1) non-metal, not smooth; (2) half-metal, half-smooth; and (3) pure metal, extremely smooth, respectively, works well in practice. 

To train our LightControlNet, we use 40K objects from the Objaverse dataset \cite{objaverse}. 
Each object is rendered from 12 views using a randomly sampled camera $C$ and lighting $L$ sampled from 6 environment maps sourced from the Internet. $L$ is also subject to random rotation and intensity scaling.
For each resulting $(L,C)$ pair, we render the conditioning image using the pre-defined materials, as well as the full-color rendering of the object using its original materials and textures.
We use the resulting 480K pairs of (conditioning images, full-color rendering) to train LightControlNet using the approach of Zhang et al.~\cite{zhang2023controlnet}. %

Once LightControlNet is trained, we can specify the desired view and lighting for any 3D mesh. We first render the conditioning image with the desired view and lighting and then pass it along with a text prompt into LightControlNet, to obtain high-quality images. These images are spatially aligned to the desired view, lit with the desired lighting, and contain detailed textures (\reffig{illum_control}).

\myparagraph{Distilling the encoder.} We improve the efficiency of SDS by distilling the image encoder in Stable Diffusion (SD)~\cite{rombach2021stablediff}, the base diffusion model in the ControlNet architecture. 
The original SD encoder consumes almost 50\% of the forward and backward time of SDS calculation, primarily in downsampling the input image. Metzer et al.~\cite{metzer2022latent} have found the image decoder can be closely approximated by per-pixel matrix multiplication. Inspired by this, we distill the encoder by removing its attention modules and training it on the COCO dataset~\cite{coco} to match the original output. This distilled encoder runs 5x faster than the original one, resulting in an approximately 2x acceleration of our text-to-texture pipeline without compromising output quality (\reftbl{ablation}). %

\begin{figure*}[t]
    \centering
    \includegraphics[width=\linewidth]{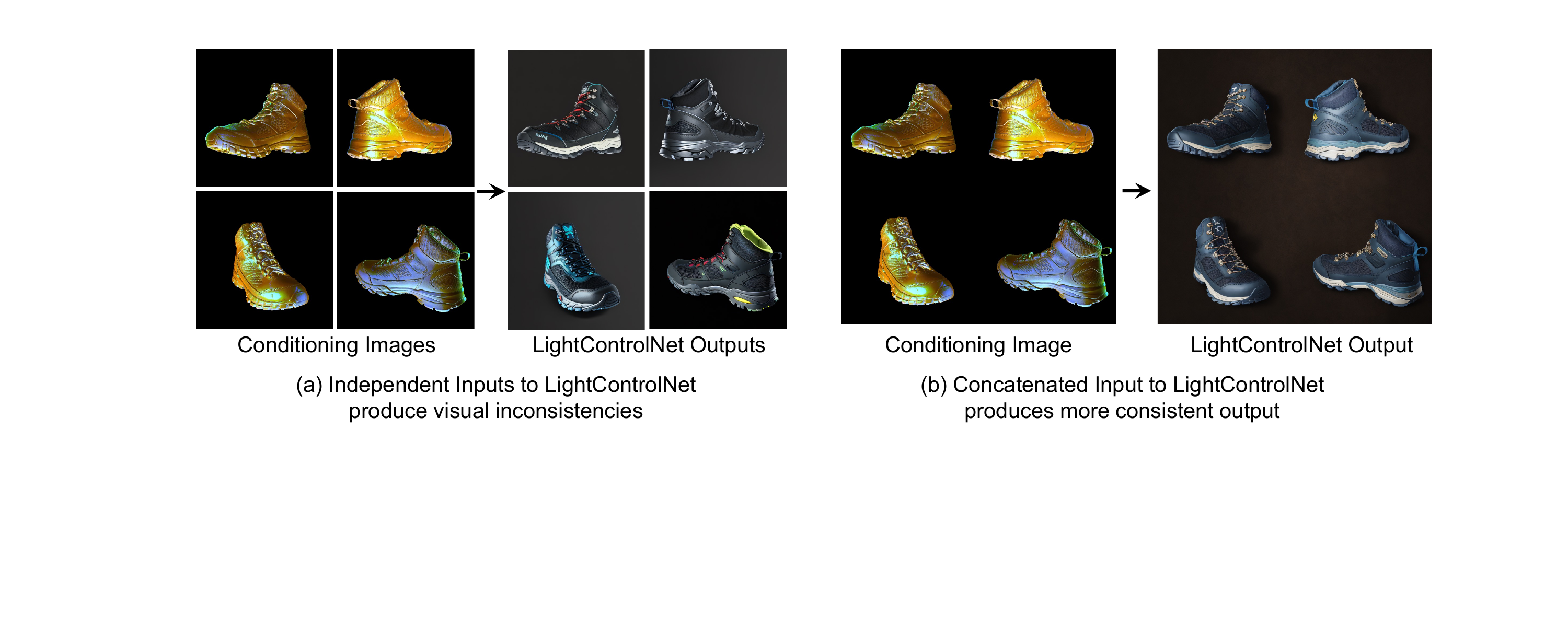}
    \caption{\textbf{Multi-view visual prompting.} (a) When we independently 
    input four canonical conditioning images to LightControlNet, it generates four very different appearances and styles even with a fixed random seed. (b) When we concatenate the four images into a 2$\times$2 grid and pass them as a single image into LightControlNet, it produces a far more consistent appearance and style. Text prompt: ``A hiking boot''. %
    }
    \lblfig{multiviewfig}
\end{figure*}

\subsection{Stage 1: Multi-view Visual Prompting}
\lblsec{stage_1}

In Stage 1, we leverage LightControlNet to synthesize high-quality 2D images for a sparse set of views of the 3D mesh. Specifically, we create conditioning images for four canonical views $C^*$ around the equator of the 3D mesh 
using a fixed lighting environment map $L^*$ sampled from a set of environment maps.
One approach to generating the complete texture for the mesh would be to apply the LightControlNet independently 
with each such conditioning image, but using the same text prompt, %
and then backprojecting the four output images to the surface of the 3D mesh. 
In practice, however, applying the LightControlNet to each view independently produces inconsistent images of varying appearance and style, even when the text prompt and random seed remain fixed (\reffig{multiviewfig}). 

We use a multi-view visual prompting approach to mitigate this multi-view inconsistency issue. 
We concatenate the conditioning images for the four canonical views into a single 2 $\times$ 2 grid and treat it as a single conditioning image. We observe that applying LightControlNet to all four views simultaneously, using this combined multi-view conditioning image, results in a far more consistent appearance and style across the views, compared to independent prompting (\reffig{multiviewfig}). 
We suspect this property arises from the presence of similar training data samples -- grid-organized sets depicting the same object -- in Stable Diffusion's training set, which is also observed in concurrent works~\cite{weber2023nerfiller, zhao2023efficientdreamer}.
Formally, we generate the conditioning image $I_{\text{cond}}(L^*, C^*)$ under a fixed canonical lighting condition $L^*$ using four canonical viewpoints $C^*$. We then apply our LightControlNet with text prompt $y$ to generate the corresponding reference image $I_{\text{ref}}$:
\begin{align*}
    I_{\text{ref}} = \text{ControlNet}(I_{\text{cond}}(L^*, C^*), y).
\end{align*}

\subsection{Stage 2: Texture Optimization} 
\lblsec{stage_2}

In Stage 2, we could directly backproject the four reference views output in Stage 1 onto the 3D mesh using the camera matrix $C$ associated with each view. 
While the resulting texture would contain some high-quality regions, it would also suffer from two problems: (1) It would contain seams and visual artifacts due to remaining inconsistencies between overlapping views, occlusions in the views that leave parts of the mesh untextured, and loss of detail when applying the backprojection transformation and resampling the views. (2) as lighting is baked into the LightControlNet's RGB images, it would also be baked into the backprojected texture, making it difficult to relight the mesh. %

To address both issues, we employ texture optimization using SDS loss. 
Specifically, we use a multi-resolution hash-grid~\cite{mueller2022instant} as our 3D texture representation. %
Given a 3D point $p \in \mathbb{R}^3$ on the mesh, our hash-grid produces a 32-dim multi-resolution feature, which is then fed to a 2-layer MLP $\Gamma$ to obtain the texture material parameters for this point. 
Similar to Fantasia3D \cite{Chen_2023fantasia3D}, these material parameters consist of metallicness $k_m \in \mathbb{R}$, roughness $k_r \in \mathbb{R}$, a bump vector $k_n \in \mathbb{R}^3$ and the base color $k_c \in \mathbb{R}^3$.
Formally, 
\begin{align*}
    (k_c, k_m, k_r, k_n) = \Gamma(\beta(p)),
\end{align*}
where $\beta$ is the multi-resolution hash encoding function. Notably, this 3D hash-grid representation can be easily converted to 2D UV texture maps, which are more friendly to downstream applications.
Given the mesh $M$, the texture $\Gamma(\beta(\cdot))$, a camera view $C$ and lighting $L$ we can use nvdiffrast~\cite{Laine2020diffrast}, a differentiable renderer $\mathcal{R}$ to produce a 2D rendering of it, $x$, as
\begin{align*}
    x = \mathcal{R}(M, \Gamma(\beta(\cdot)), L, C).
\end{align*}
More details about the rendering equation are in the appendix.
Since the mesh geometry is fixed, we omit $M$ in the remainder of the paper. 

Recall that the optimization approach of DreamFusion~\cite{poole2022dreamfusion} randomly samples camera views $C$, generates an image for $C$ using diffusion model $\phi$, and supervises the optimization using the SDS loss. We extend this optimization in two ways. First, we use four fixed reference images $I_\text{ref}$ with their canonical views $C^*$ and lighting $L^*$ to guide the texture optimization through a reconstruction loss: %

\begin{align*}
    \mathcal{L}_{\text{recon}}   = %
    ||I_{\text{ref}}-\mathcal{R}(\Gamma(\beta(\cdot)), L^*, C^*)||_2 + \mathcal{L}_{\text{perceptual}}(I_{\text{ref}}, \mathcal{R}(\Gamma(\beta(\cdot)), L^*, C^*)), 
\end{align*}
where  both L2 loss and perceptual loss~\cite{johnson2016perceptual} are used. 
For a non-canonical view $C$, we sample a random lighting $L$ and use the SDS loss to supervise the optimization, but with our LightControlNet as the diffusion model $\phi_\text{LCN}$, so  
\begin{align*}
     \nabla_{\Gamma, \beta} \mathcal{L}_{\text{SDS}}(\phi_\text{LCN}, x)  =  \mathbb{E}_{t,\epsilon}\left[ w(t) (\hat{\epsilon}_{\phi_\text{LCN}}(x_t; y, t, I_{\text{cond}}(L,C)) - \epsilon) \frac{\partial x}{\partial \Gamma(\beta(\cdot))} \right],
\end{align*}
where $x = \mathcal{R}(\Gamma(\beta(\cdot)), L, C)$ and $w(t)$ is the weight. %

Finally, we employ a material smoothness regularizer on every iteration to enforce smooth base colors, using the approach of nvdiffrec~\cite{Munkberg_2022_nvdiffrec}. For a surface point $p$ with base color $k_c(p)$, the smoothness regularizer is defined as
\begin{align*}
    \mathcal{L}_{\text{reg}} = \sum_{p \in S} |k_c(p) - k_c(p+\epsilon)|,
\end{align*}
where $S$ denotes the object surface and $\epsilon$ is a small random 3D perturbation. We use $\lambda_{\text{recon}}=1000$ and $\lambda_{\text{reg}}=10$ to reweight the loss $ \mathcal{L}_{\text{recon}} $ and $ \mathcal{L}_{\text{reg}} $.

\myparagraph{Scheduling the optimization.} We warm up the optimization by rendering the four canonical views and applying $\mathcal{L}_{\text{recon}}$ for 50 iterations. 
We then add in iterations using $\mathcal{L}_{\text{SDS}}$ and optimize over 
randomly chosen camera views and randomly selected lighting
from a pre-defined set of environmental lighting maps. Specifically we alternate iterations between using $\mathcal{L}_{\text{SDS}}$ and 
$\mathcal{L}_{\text{recon}}$. In addition, for a quarter of the SDS iterations, we use the canonical views rather than randomly selecting the views. This ensures that the resulting texture does not overfit to the reference images corresponding to the canonical views. 
The warm-up iterations capture the large-scale structure of our texture and allow us to use relatively small noise levels ($t \leq 0.1$) in the SDS optimization. We sample the noise 
following a linearly decreasing schedule \cite{huang2023dreamtime} with $t_{\text{max}}=0.1$ and $t_{\text{min}}=0.02$.
We also adjust the conditioning strength $s$ of our LightControlNet in $\mathcal{L}_{\text{SDS}}$ linearly from $1$ to $0$ over these iterations so that LightControlNet is only lightly applied by the end of the optimization. We also experimented with a recent variant Variational Score Distillation~\cite{wang2023prolificdreamer}, but did not observe notable improvement. 
We have experimentally found that we obtain high-quality textures after 400 total iterations of this optimization and this is far fewer iterations than other SDS-based texture generation techniques such as Fantasia3D~\cite{Chen_2023fantasia3D} which requires 5000 iterations. %

\myparagraph{Faster pipeline without relightability.} Our two-stage pipeline is also compatible with off-the-shelf depth ControlNet and Stable Diffusion~\cite{rombach2021stablediff} as the backbone replacement of LightControlNet. Specifically, we can replace the LightControlNet in Stage 1 with a depth ControlNet that uses a depth rendering of the mesh as the conditioning image, and uses Stable Diffusion based SDS in Stage 2. In scenarios where texture relightability is not required, this variant offers an additional 2$\times$ speed-up (as shown in \reftbl{quan_eval}), since it eliminates the additional computation required by LightControlNet forward pass in the SDS optimization.

\section{Experiments}
\lblsec{experiment}

\begin{figure*}[t]
    \centering
    \includegraphics[width=\linewidth]{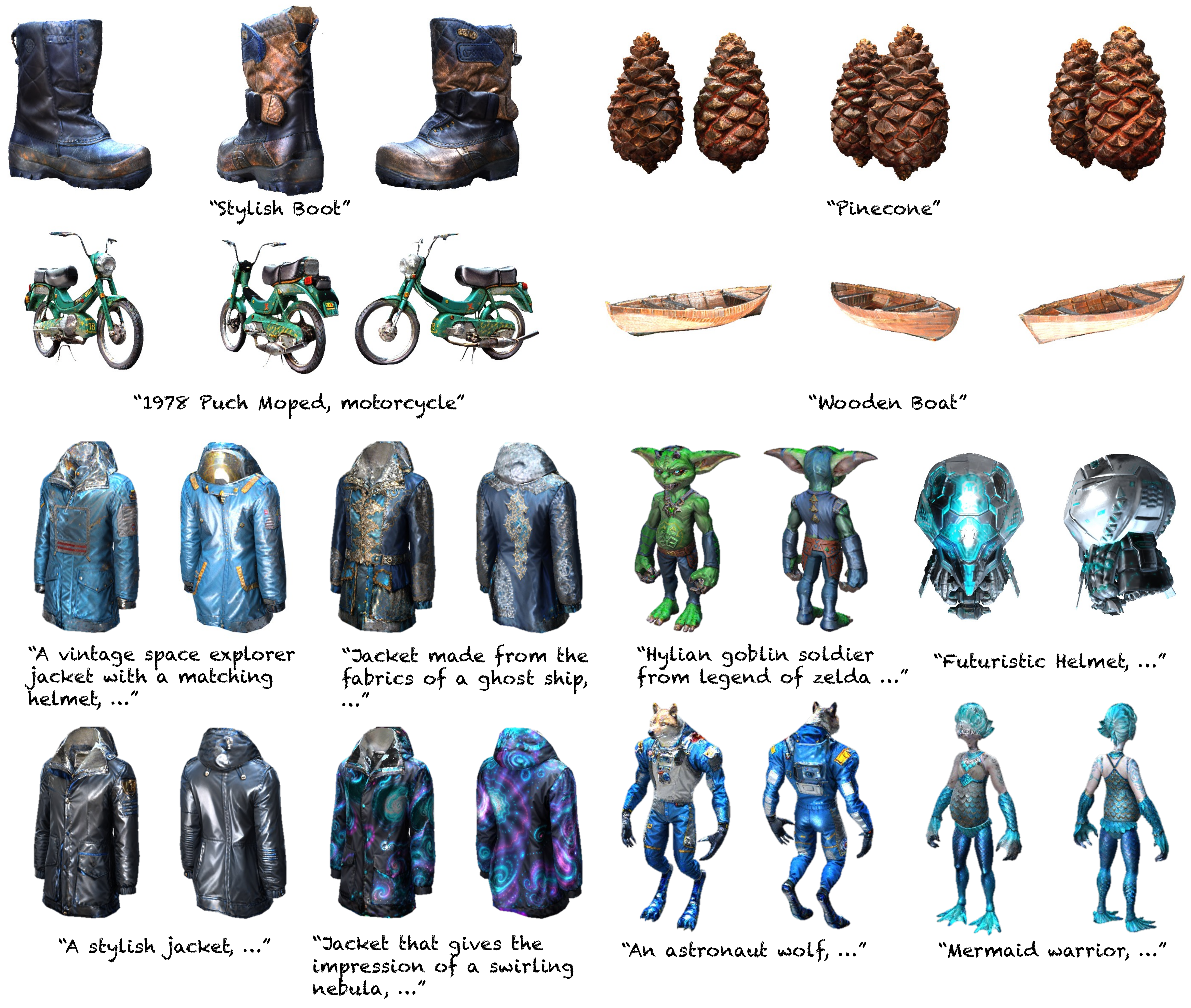}
    \caption{{\bf Sample results} from our method applied to Objaverse test meshes (top half) and 3D game assets (bottom half). To illustrate the efficacy of our relightable textures, for each textured mesh, we fix the environment lighting and render the mesh under different rotations. As shown above, our method is able to generate textures that are not only highly detailed, but also relightable with realistic lighting effects.} 
    \lblfig{result}
\end{figure*}

In this section, we present comprehensive experiments to evaluate the efficacy of our proposed method for relightable, text-based mesh texturing. We perform both qualitative and quantitative comparisons with existing baselines, along with an ablation study on the significance of each of our major components.

\myparagraph{Dataset.} As illustrated in \reffig{pipeline}, we employ Objaverse \cite{objaverse} to render paired data to train our LightControlNet. Objaverse consists of approximately 800k objects, of which we use the names and tags as their text descriptions. We filter out objects with low CLIP similarity \cite{radford2021clip} to their text descriptions and select around 40k as our training set.
To evaluate baselines and our method, we hold out 70 random meshes from Objaverse \cite{objaverse} as the test set. We additionally gather 22 mesh assets from 3D online games with 5 prompts each to assess our method. %

\myparagraph{Baselines.} We compare our approach with existing mesh texturing methods. Specifically, Latent-Paint \cite{metzer2022latent} employs SDS loss in latent space for texture generation. Text2tex \cite{chen2023text2tex} progressively produces 2D views from chosen viewpoints, followed by an inverse projection to lift them to 3D. TEXTure \cite{richardson2023texture} utilizes a similar lifting approach but supplements it with a swift SDS optimization post-lifting. Beyond these texture generation methods, text-to-3D approaches serve as additional baselines, given that texture is a component of 3D generation. Notably, we choose Fantasia3D \cite{Chen_2023fantasia3D} as a baseline, the first to use a material-based representation for textures in text-to-3D processing.

\begin{figure*}[t]
    \centering
    \includegraphics[width=\linewidth]{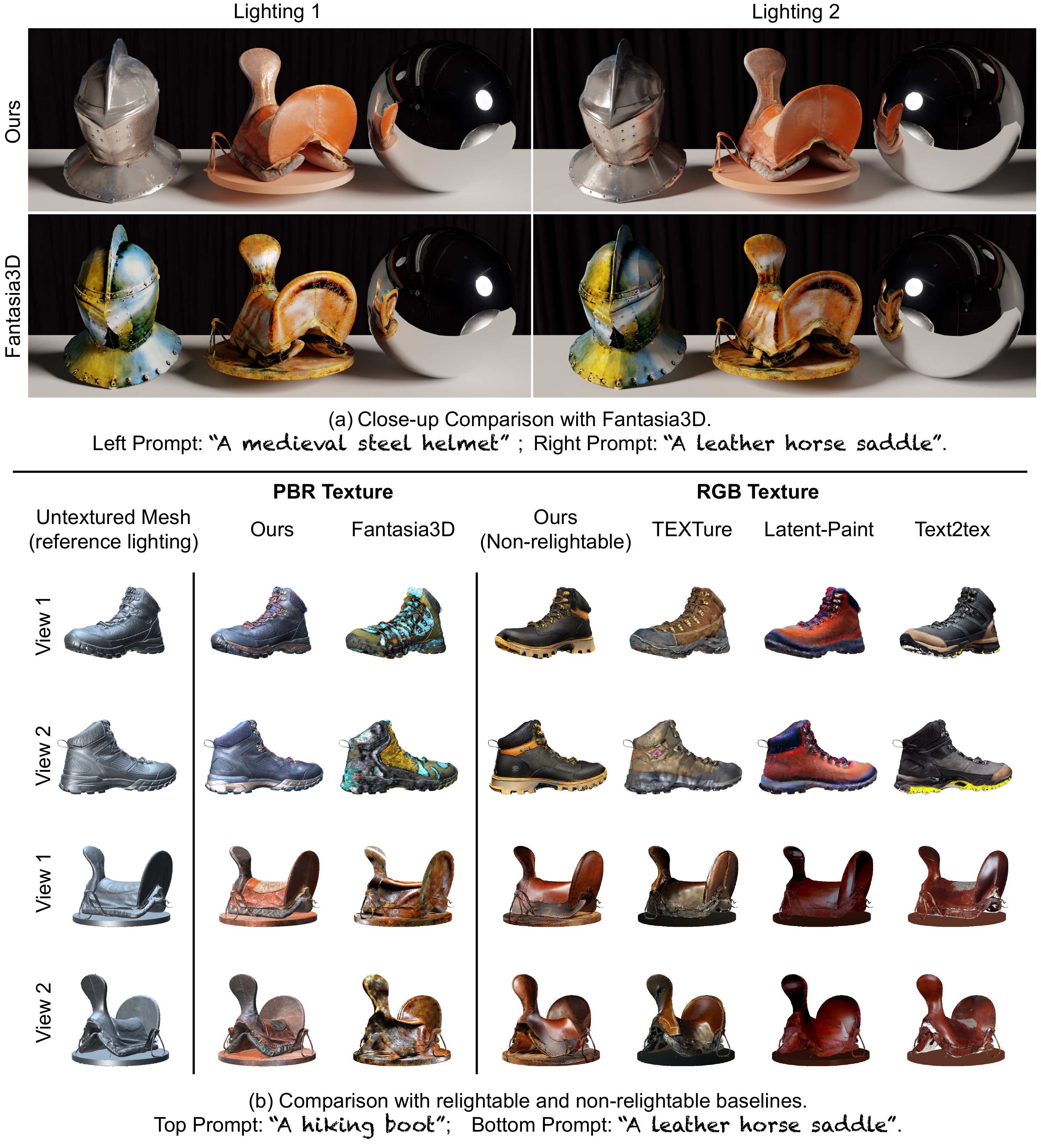}
    \caption{\textbf{Qualitative analysis.} 
    (a) We compare with Fantasia3D \cite{Chen_2023fantasia3D} that also attempts to generate Physically Based Rendering (PBR) texture. However, their results often exhibit baked-in lighting, leading to artifacts when put under varied lighting.
    (b)
    We also compare with other baselines that only generate non-relightable (RGB) texture. For non-relightable texture generation, we can replace our LightControlNet with depth ControlNet and generate textures with a shorter runtime. More details are in \reftbl{quan_eval}.} 
    \lblfig{comparison}
\end{figure*}

\begin{table}[t]
    \centering

        \caption{\textbf{Quantitative Evaluation.} We test our methods and baselines on 70 test objects from Objaverse \cite{objaverse} and 22 objects curated from 3D game assets. With depth ControlNet, our method yields superior results to all baselines while being three times as fast as the fastest baseline. Using LightControlNet (Ours) within our model improves the lighting disentanglement while maintaining comparable image quality. }
    \setlength{\tabcolsep}{10pt}
    \begin{tabular}{lccccc}
    \toprule
    & \multicolumn{2}{c}{Objaverse test set} & \multicolumn{2}{c}{Game Asset} & \multirow{2}{*}{Runtime $\downarrow$}  \\
    \cmidrule(rl){2-3} \cmidrule(rl){4-5} 
       & FID $\downarrow$  & KID $\downarrow$ & FID $\downarrow$  & KID $\downarrow$ &    \\
     & & ($\times 10^{-3}$) &  & ($\times 10^{-3}$) &  (mins) \\
    \midrule
    Latent-Paint \cite{metzer2022latent}  & 73.65 & 7.26 & 204.43 & 9.25  & 10 \\
    Fantasia3D \cite{Chen_2023fantasia3D} & 120.32 & 8.34 & 164.32 &  9.34 & 30 \\
    TEXTure \cite{richardson2023texture} & 71.64 & 5.43 & 103.49 & 5.64  & 6 \\
    Text2tex \cite{chen2023text2tex} & 95.59 & 4.71 & 119.98 & 5.21 & 15 \\
    \midrule
    Ours (w/ depth) & \textbf{60.49} & 3.96 & 85.92 & 3.87 &  \textbf{2} \\
    
    Ours  & 62.67 & \textbf{2.69} & \textbf{83.32} & \textbf{3.34} & 4 \\
    \bottomrule
    \end{tabular}

    \lbltbl{quan_eval}
\end{table}

\begin{table}
    \centering
    
    \caption{\textbf{User study.}  We conduct a user preference study to evaluate %
    (1) result realism, (2) texture consistency with input text, and (3) plausibility under varied lighting. 
    Participants consistently prefer our results over all baselines in these respects.%
    }
     \setlength{\tabcolsep}{9pt}
    \begin{tabular}{lccc}
    \toprule
    \multirow{2}{*}{Preferred Percentage} & \multicolumn{3}{c}{Objaverse test set} \\
    \cmidrule{2-4}
       &  Realistic & Consistent with text & Relightable \\
    \midrule
    Ours v.s. Latent-Paint \cite{metzer2022latent}  & $92.6\%$ & $74.5\%$  & $84.3\%$ \\
    Ours v.s. Fantasia3D \cite{Chen_2023fantasia3D} & $81.9\%$ & $67.6\%$ & $74.3\%$ \\
    Ours v.s. TEXTure \cite{richardson2023texture} & $70.8\%$ & $57.3\%$ & $87.1\%$ \\
    Ours v.s. Text2tex \cite{chen2023text2tex} & $75.4\%$ & $61.6\%$ & $88.6\%$ \\
    \bottomrule
    \end{tabular}
    \lbltbl{user_study}
\end{table}

\myparagraph{Quantative evaluation.} In \reftbl{quan_eval}, we compare our method with the baselines on the Objaverse \cite{objaverse} test set. For each method, we generate 16 views and evaluate Frechet Inception Distance (FID) \cite{heusel2017fid, parmar2021cleanfid} and Kernel Inception Distance (KID) \cite{binkowski2018kid} compared with ground-truth renderings. Two variations of our method are assessed. Both variants use ourr two-stage pipeline, and the first employs a standard depth ControlNet, while the second uses our proposed LightControlNet. Our method outperforms the baselines in both quality and runtime.

\myparagraph{Qualitative analysis.} In \reffig{result}, our method can generate highly detailed textures that can be rendered properly with the environment lighting across various meshes. We also visually compare our method and the baselines
in \reffig{comparison}. Our method produces textures with higher visual fidelity for both the relightable and non-relightable variants. In particular, when compared with Fantasia3D \cite{Chen_2023fantasia3D}, a recent work that also aims to generate material-based texture, our results not only have superior visual quality, but also disentangle the lighting more successfully. 

\myparagraph{User study.} To further evaluate the texture quality quantitatively, we conduct a user study comparing our results with each of the baselines on the Objaverse test set in \reftbl{user_study}. We asked 30 participants to evaluate (1) the realism of the results, (2) the consistency of the generated texture with the input text, and (3) the plausibility of the results when placed under varying lighting conditions. Each result is presented in the form of 360-degree rotation to display full texture details. The reference lighting is provided alongside when participants evaluate (3). Across all three aspects, participants consistently prefer our method.

\begin{table}[!t]
    \centering
    \caption{\textbf{Ablation study on algorithmic components.} We analyze the role of our distilled encoder (1st row) and multi-view visual prompting (2nd row). Replacing the distilled encoder with the original encoder doubles the running time without a noticeable improvement. When removing the multi-view visual prompting for initial generation, the system requires 2,000 iterations (5x compared to our 400 iterations) to produce reasonable results, which produces slightly worse texture quality.}
    \setlength{\tabcolsep}{10pt}
    \begin{tabular}{lccc}
    \toprule
    Objaverse test set   & FID $\downarrow$  & KID ($\times 10^{-3}$) $\downarrow$ & Runtime $\downarrow$ (mins)  \\
    \midrule
    Ours (w/o dist. enc.) & \textbf{60.34} & 2.84 & 8 \\
    Ours (w/o m.v.v.p)  & 74.23 & 3.54 & 19 \\
    \midrule
    Ours & 62.67 & \textbf{2.69} &  \textbf{4} \\
    \bottomrule
    \end{tabular}
    \lbltbl{ablation}
\end{table}

\begin{table}[!t]
    \centering
    \caption{\textbf{Ablation study on material bases.} We verify the impact of the material bases in rendering conditioning images. %
    Omitting any one of these degrades quality.}
    \setlength{\tabcolsep}{10pt}
    \begin{tabular}{ccccc}
    \toprule
     \multicolumn{3}{c}{Material Basis} \\
    \cmidrule{1-3}
    non-metal, & half-metal,  & pure metal, &  FID $\downarrow$  & KID ($\times 10^{-3}$) $\downarrow$  \\
      not smooth & half-smooth &  smooth  \\
    \midrule
         \checkmark  & \checkmark  & \checkmark & \textbf{62.67} & \textbf{2.69} \\
    \midrule
     &   \checkmark  & \checkmark & 66.34 & 3.11  \\
     \checkmark  &   & \checkmark & 64.32 & 3.42  \\
    \checkmark  & \checkmark  &  & 67.43 & 4.12  \\
      & \checkmark  &  & 72.13 & 4.53 \\
    \bottomrule
    \end{tabular}
    
    \lbltbl{ablation_material}
\end{table}

\begin{table}[!t]
    \centering
    \caption{\textbf{Ablation study on canonical view selection in \refsec{stage_1}.} 
     Using only front and back views provides insufficient supervision while adding top and bottom views worsens quality. This likely stems from pre-trained 2D diffusion models struggling with top and bottom views. Additionally, stacking more views reduces each view’s resolution, leading to poorer initialization for Stage 2.
    }

    \setlength{\tabcolsep}{10pt}
    \begin{tabular}{lcc}
    \toprule
    Num. of canonical views   & FID $\downarrow$  & KID ($\times 10^{-3}$) $\downarrow$ \\
    \midrule
    2 views (front, back) & 67.43 & 3.47  \\
    4 views (\textbf{Ours:} front, back, left, right) & \textbf{62.67} & \textbf{2.69} \\
    6 views (front, back, left, right, top, bottom) & 70.14 & 3.72  \\
    \bottomrule
    \end{tabular}
    \lbltbl{ablation_view}
\end{table}

\myparagraph{Ablation study.}  
We perform an ablation analysis on different aspects of our method in \reftbl{ablation}.
When replacing our distilled encoder with the original SD encoder, performance is twice as slow without noticeably superior quality. On the other hand, without the multi-view visual prompting for the initial generation, the system requires 2000 iterations (a 5x slowdown compared to our 400 iterations) to produce reasonable results while still leading to slightly worse texture quality. 
In \refsec{LightControlNet}, we render a conditioning image using three pre-defined materials to encompass a broad range of feasible effects. \reftbl{ablation_material} shows omitting any one of these bases degrades quality. \reftbl{ablation_view} evaluates our selection of four canonical views in \refsec{stage_1}. Relying on only the front and back views provides insufficient supervision. Interestingly, incorporating top and bottom views degrades the performance. We hypothesize that this is likely due to the limitation of 2D diffusion model backbones in reliably generating top and bottom views. Furthermore, stacking more views within a single image results in a decreased resolution for each view, given the fixed resolution of the multi-view image.

\section{Discussion}

We proposed an automated texturing technique based on user-provided prompts. Our method employs an illumination-aware 2D diffusion model (LightControlNet) and an improved optimization process based on the SDS loss. Our approach is substantially faster than previous methods while yielding high-fidelity textures with illumination disentangled from surface reflectance/albedo. We demonstrated the efficacy of our method through quantitative and qualitative evaluation on the Objaverse dataset and meshes curated from game assets. 

\myparagraph{Limitations.}
Our approach still poses a few limitations: (1) Baked-in lighting can still be found in certain cases, especially for meshes that are outside of the training data distribution; (2) The generated material maps are sometimes not fully disentangled and interpretable as metallicness, roughness, etc.

\clearpage
\section*{Acknowledgements}
We thank Benjamin Akrish, Victor Zordan, Dmitry Trifonov, Derek Liu, Sheng-Yu Wang, Gaurav Parmer, Ruihan Gao, Nupur Kumari, and Sean Liu for their discussion and help. This work was done when KD was an intern at Roblox. The project is partly supported by Roblox. JYZ is partly supported by the Packard Fellowship. The Microsoft Research PhD Fellowship supports KD.

\bibliographystyle{splncs04}
\bibliography{main}

\clearpage
\appendix

\section{Implementation Details and Additional Results}
\myparagraph{Distilled Encoder.}
In Section 4.1 of the main paper, we improve the efficiency of LightControlNet by distilling the image encoder in Stable Diffusion \cite{rombach2021stablediff}. %
We profile the running time of our distilled and original encoder in the following table.

\begin{table}[!h]
    \centering
    \begin{tabular}{cccc}
    \toprule
    Time (ms)    & Forward  & Backward & Total \\
    \midrule
    Original     & 113 & 569 & 682 \\
    Distilled     & \textbf{42} & \textbf{81} & \textbf{123} ($5.5 \times$) \\
    \bottomrule
    \end{tabular}
    \caption{We profile the forward and backward pass of our distilled and original encoder in Stable Diffusion \cite{rombach2021stablediff} on an A100 GPU. Our distilled encoder runs more than $5 \times$ faster than the original one for a single forward and backward pass.}
    \lbltbl{profile_encoder}
\end{table}

\myparagraph{Hyper-parameters.} We provide the hyper-parameters used by our pipeline in Section 4.3 of the main paper. 
We use a batch of 4, a learning rate of 0.01 for optimization, and a CFG scale of 50 in Score Distillation Sampling loss. In Section 4.1 and Section 4.2 (main paper), we set 3 pre-defined materials to generate a conditioning image. The specific material parameters are (1) non-metal, non-smooth: $k_m=0, k_r=1$; (2) half metal, half smooth: $k_m=0.5, k_r=0.5$; (3) pure metal, extremely smooth: $k_m=1, k_r=0$. The color $k_c$ is always set to $(1,1,1)$.  We train our LightControlNet for 20,000 iterations with a batch size of 16.

\myparagraph{Base Models.} We use stable diffusion v1.5 (\textit{SD1.5}) as our base model for the experiments in Tables 1, 2, and 3 of the main paper.
Our pipeline is also compatible with other base models fine-tuned from \textit{SD1.5}. For example, we have also used \href{https://civitai.com/models/4384/dreamshaper}{\textit{Dreamshaper}}, a community fine-tuned checkpoint of SD1.5, to generate a variety of captivating textures. %
We include some of the results in Figure 6 (main paper) and \reffig{result_supp}. Specifically, the results of the jackets, goblins, fishmen, and wolves are obtained using \textit{Dreamshaper}.

\myparagraph{Environmental Light Maps.} In Section 4, we use randomly rotated environmental light maps to represent different lighting conditions. Specifically, we download 6 HDRI light maps from \href{https://polyhaven.com/a/studio_small_01}{polyhaven}. These HDRI maps are captured in a studio environment. We show these light maps in \reffig{env_map}. We also study the effect of different lighting by using various environment maps in training LightControlNet. We contrast the indoor set used in the main paper (FID: 62.67, KID: 2.69) with an outdoor set of 6 maps featuring brighter ambient light (FID: 63.23, KID: 2.66), and another set of 6 maps with less directional lighting (FID: 70.14, KID: 3.72). The results suggest directional lighting is important in accurately modeling lighting effects with LightControlNet, whereas ambient light intensity minimally impacts quality. 

\begin{figure}[t]
    \centering
    \includegraphics[width=\linewidth]{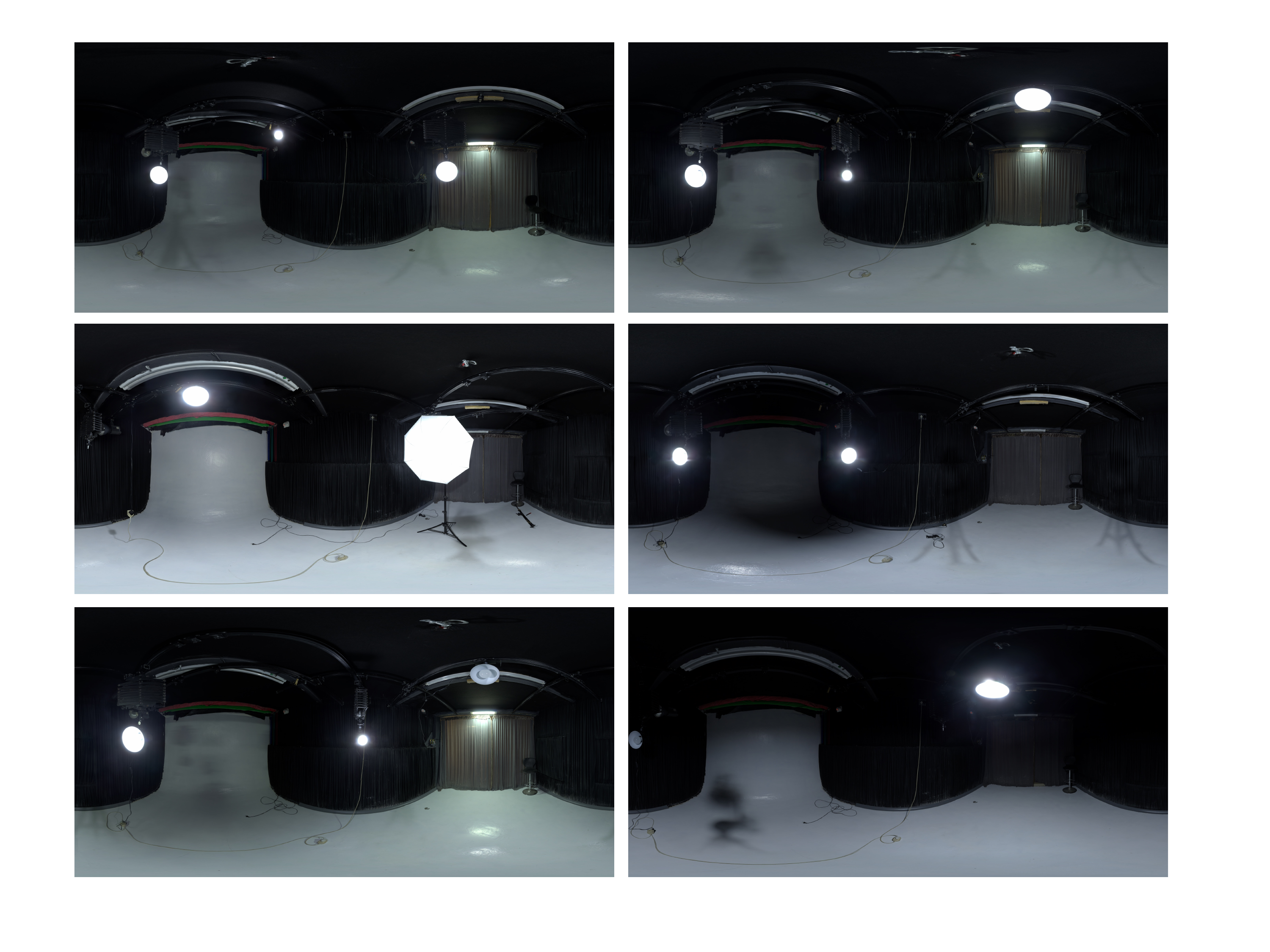}
    \caption{\textbf{Environmental Light Maps.} We download 6 HDRI maps from \href{https://polyhaven.com/a/studio_small_01}{polyhaven} to represent different different lighting conditions. For random lighting samples, we select one map from them and apply a random rotation.} 
    \lblfig{env_map}
\end{figure}

\myparagraph{BRDF Model and Rendering Equation.} As described in Section 4.3 of the main paper, our material model~\cite{walter2007microfacet} consists of metallicness $k_m \in \mathbb{R}$, roughness $k_r \in \mathbb{R}$, a bump vector $k_n \in \mathbb{R}^3$ which is a perturbation of surface normal in tangent space, and the base color $k_c \in \mathbb{R}^3$. We show an example of our generated material maps in \reffig{material}. The general rendering equation is:

\begin{equation*}
L(p, \omega)=\int_{\Omega} L_i\left(p, \omega_i\right) f\left(p, \omega_i, \omega\right)\left(\omega_i \cdot n_p\right) \mathrm{d} \omega_i,
\end{equation*}
where $L$ is the rendered pixel color at the surface point $p$ from the direction $\omega$, $\Omega$ denotes a
hemisphere with the surface normal $n_p$ at $p$. $L_i$ is the incident light represented by an environment map, and $f$ is the BRDF function determined by the material parameters $(k_m, k_r, k_n, k_c)$. $L$ can be calculated as the summation of diffuse intensity $L_d$ and specular intensity $L_s$ as follows:

\begin{align*}
& L(p, \omega)  =L_d(p)+L_s(p, \omega), \\
& L_d(p) =k_c(1-k_m) \int_{\Omega} L_i\left(p, \omega_i\right)\left(\omega_i \cdot n_p\right) \mathrm{d} \omega_i, \\
& L_s(p, \omega) = \\ 
& \int_{\Omega} \frac{D(n_p) F(\omega_i,\omega,n_p) G(\omega_i, \omega, n_p)}{4\left(\omega \cdot n_p\right)\left(\omega_i \cdot n_p\right)} L_i\left(p, \omega_i\right)\left(\omega_i \cdot n_p\right) \mathrm{d} \omega_i,
\end{align*}
where $F$, $G$, and $D$ are functions representing the Fresnel term, the geometric attenuation, and the GGX normal distribution \cite{walter2007microfacet}, respectively.
Following Nvdiffrec \cite{Munkberg_2022_nvdiffrec} and Fantasia3D \cite{Chen_2023fantasia3D}, the hemisphere integration can be calculated using the split-sum method.

\myparagraph{Rendering Engine.} We use nvdiffrast \cite{Laine2020diffrast}, a differentiable renderer, as our rendering engine in both of our training and testing phases. Notably, our exported material maps are also directly compatible with widely used rendering applications, such as Blender. This compatibility is demonstrated in the main paper, where Figure 1 and the top half of Figure 7 are rendered using Blender, showcasing the practical application of our generated textures.

\begin{figure}[t]
    \centering
    \includegraphics[width=\linewidth]{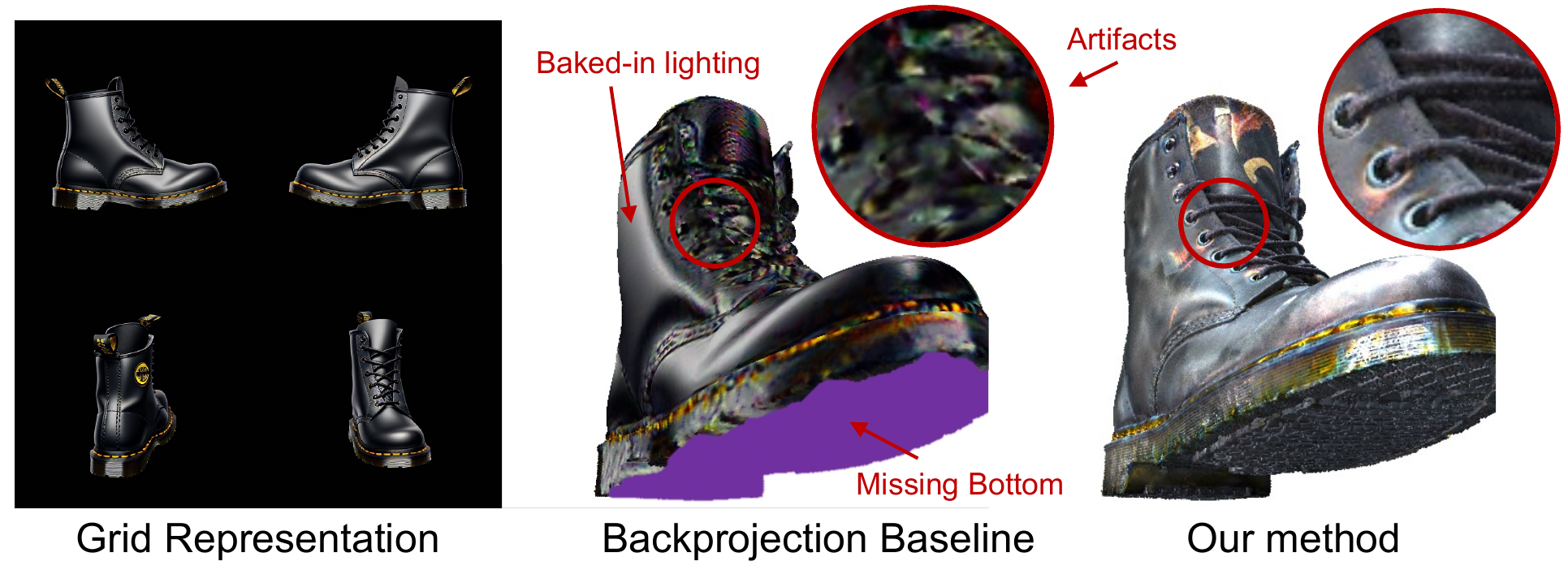}
    \caption{\textbf{Backprojection Baseline.} Directly backprojecting the grid representation onto 3D mesh leads to baked-in lighting, stitching artifacts, and missing areas of texture. }
    \lblfig{backprojection}
\end{figure}

\begin{figure*}
    \centering
    \vspace{-1em}
    \includegraphics[width=\linewidth]{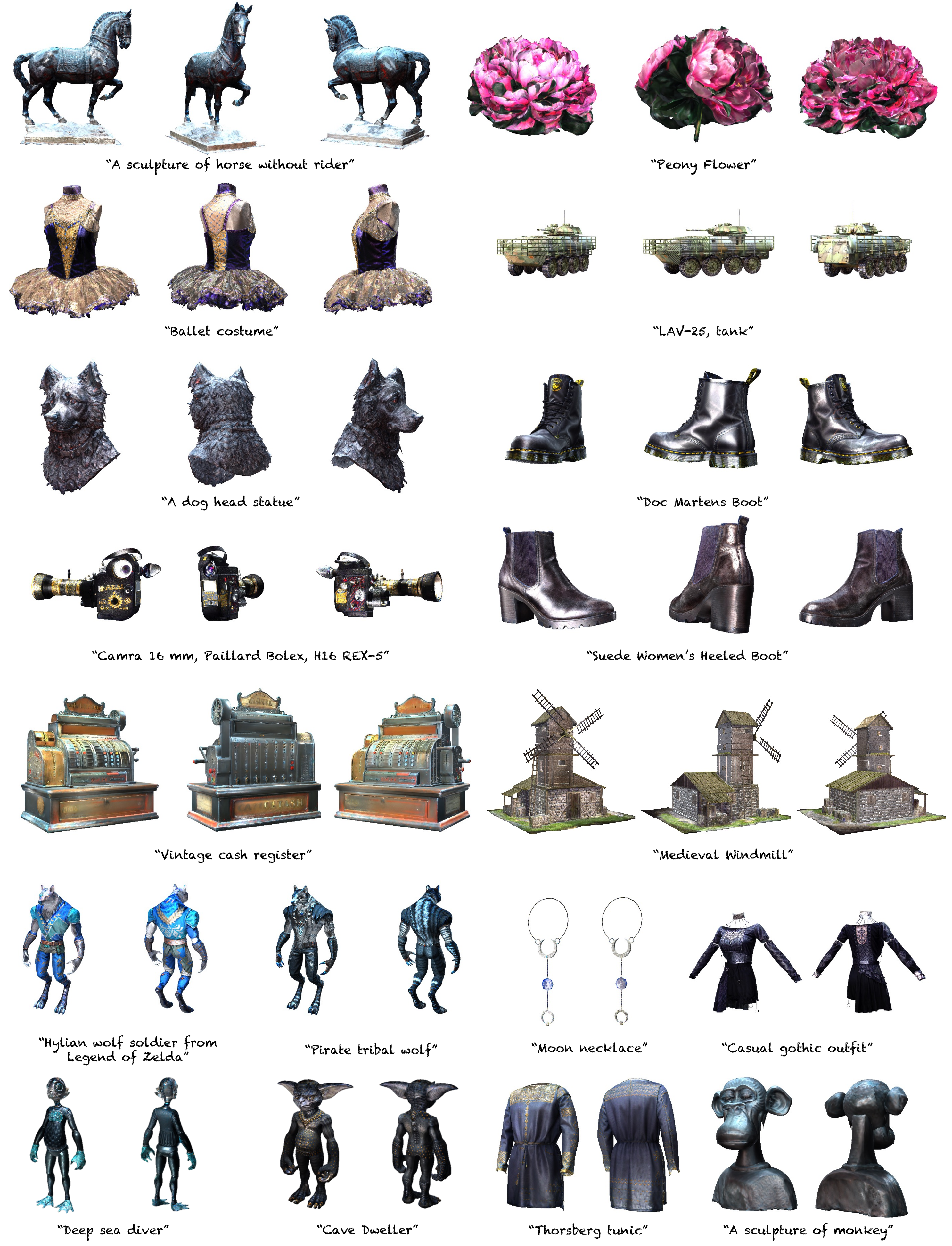}
    \caption{\textbf{Additional Results.} We rotate the objects to show different views while fixing the lighting condition.} 
    \lblfig{result_supp}
\end{figure*}

\begin{figure*}
    \centering
    \vspace{-1em}
    \includegraphics[width=.95\linewidth]{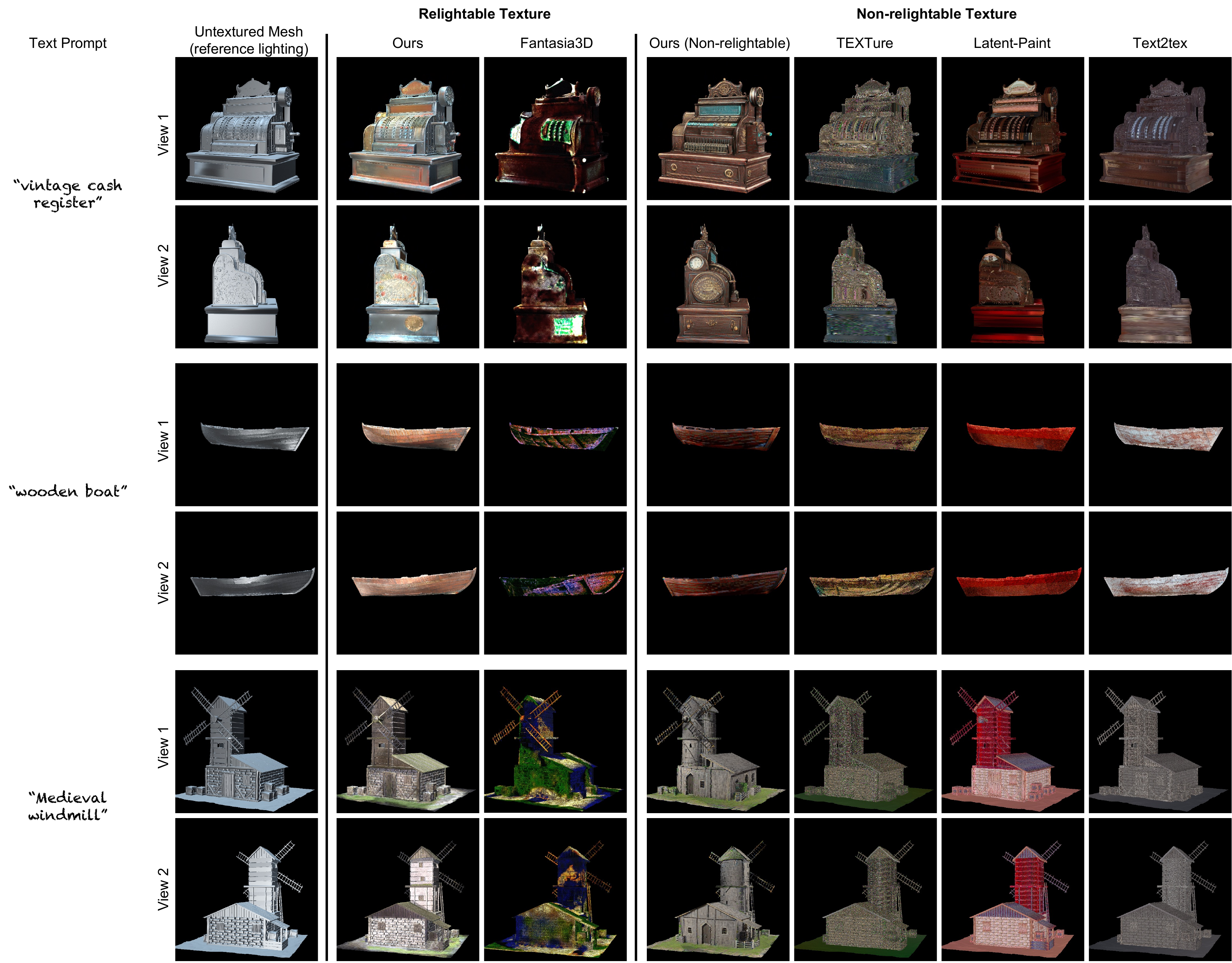}
    \caption{\textbf{Additional Qualitative Analysis.} We present additional comparisons with the baselines. To evaluate the quality of relightable textures, our results and those from Fantasia3D are placed under identical lighting conditions. An untextured mesh is also rendered under this condition to denote the reference lighting direction. Our results adhere to the lighting conditions and generally outperform all existing baselines in quality.} 
    \lblfig{comparison_supp}
\end{figure*}

\myparagraph{Backprojection Baseline.} We mention an alternative baseline by directly backprojecting the sparse views in Section 4 of the main paper. We compare this baseline with our method in \reffig{backprojection}.

\myparagraph{Additional Results.} We show additional results in Figures~\ref{fig:result_supp} and \ref{fig:comparison_supp}. 
We also show an example of our generated material maps in \reffig{material}.

\begin{figure}[t]
    \centering
    \includegraphics[width=\linewidth]{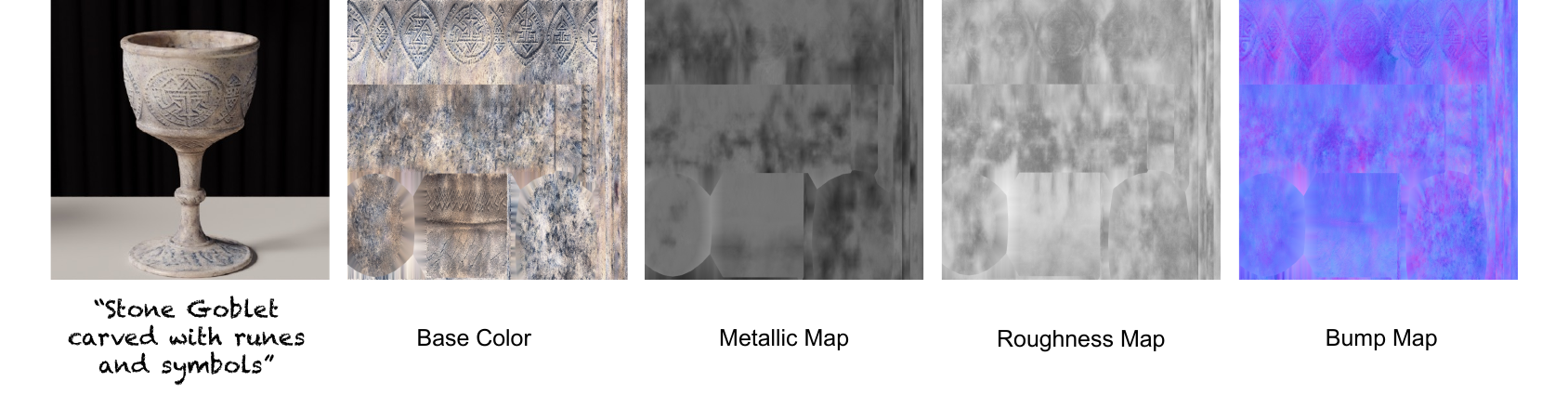}
    \caption{We show an example of our \textbf{generated material maps}.}
    \lblfig{material}
\end{figure}

\section{Additional Related Works}

In Section 2 of the main paper, we discuss prior works on \textbf{3D texture generation}. Another excellent recent work, TexFusion \cite{cao2023texfusion}, aligns with the research using depth-conditioned diffusion models for texture creation. They introduce a Sequential Interlaced Multiview Sampler, which performs denoising steps across multiple camera viewpoints and integrates the outcomes into a unified latent texture map. The execution time for TexFusion is comparable to that of TEXTure\cite{richardson2023texture}, with the comprehensive approach taking approximately 6 minutes, whereas a quicker variant takes about 2.2 minutes but yields lower-quality results. Similar to both TEXTure\cite{richardson2023texture} and Text2tex\cite{chen2023text2tex}, TexFusion's textures are entangled with lighting effects. The non-relightable textures make it challenging to adapt to various lighting conditions. On the other hand, our generated texture can be relit properly in different lighting environments with a comparable runtime (4 mins). As their code is unavailable, we cannot preform a direct comparison. However, we will be happy to compare our method with this baseline once their code becomes available.

\section{Societal Impact}

Our method in text-based mesh texturing will enable many applications in 3D content creation. 
First, our method drastically reduces the time and expertise required for texture authoring, making 3D content generation accessible to a broader audience.
Additionally, our faster inference time and improved result quality also reduce computational costs and energy consumption.  
However, our method might also be misused to generate fake content for misinformation. Nevertheless, we believe humans can currently distinguish our synthesized objects from photo captures of real objects.

\end{document}